\def\upd{{\rm d}}

\documentclass[pre,12pt]{revtex4}
\usepackage{graphicx}

\newcommand{\be}{\begin{eqnarray}}
\newcommand{\ee}{\end{eqnarray}}

\begin{document}

\title{Path integral approach to random motion with nonlinear friction}

\author{A. Baule$^1$, E. G. D. Cohen$^1$, and H. Touchette$^2$}

\affiliation{                    
$^1$The Rockefeller University, 1230 York Avenue, New York, NY 10065, USA\\
$^2$School of Mathematical Sciences, Queen Mary University of London, London E1 4NS, UK
}

\begin{abstract}

Using a path integral approach, we derive an analytical solution of a nonlinear and singular Langevin equation, which has been introduced previously by P.-G. de Gennes as a simple phenomenological model for the \textit{stick-slip} motion of a solid object on a vibrating horizontal surface. We show that the optimal (or most probable) paths of this model can be divided into two classes of paths, which correspond physically to a sliding or slip motion, where the object moves with a non-zero velocity over the underlying surface, and a stick-slip motion, where the object is stuck to the surface for a finite time. These two kinds of basic motions underlie the behavior of many more complicated systems with solid/solid friction and appear naturally in de Gennes' model in the path integral framework.

\end{abstract}

\date{\today}

\maketitle

\section{Introduction}

We study an old but still only very partially understood problem: the dynamics of a solid object moving over a solid surface. In practice this is a very complicated and as yet unsolved problem, although there is a wealth of experiments, since the general problem is very old and ubiquitous in nature, ranging from geology to physics and biology. The basic difficulty lies in the very complex nature and behavior of the solid/solid interface, which leads to a complicated stick-slip motion of the object \cite{Persson}. 

Following P.-G. de Gennes we study in detail one of the simplest phenomenological models, far from those of most practical interest, but as a starting point to develop a new theoretical approach to describe basic aspects of the above mentioned problem. Ignoring all details of the solid/solid interfacial layer, de Gennes proposed a simple Langevin equation for the velocity $v(t)$ of a solid object of mass $m$ on a horizontal vibrating surface \cite{DeGennes05,Buguin06}:
\begin{eqnarray}
\label{deGennes}
m\dot{v}(t)+\alpha v(t)+\sigma[v(t)]\Delta_F=\xi(t).
\end{eqnarray}
In this equation two kinds of friction between the object and the surface, over which it moves, appear: a) a \textit{dynamic} friction (sometimes called kinetic friction), which is taken proportional to $v$ as in the Stokes friction term in fluids and characterized by the dynamical friction coefficient $\alpha$; b) a \textit{static} friction (sometimes called dry friction), which is given by the $\sigma(v)\Delta_F$ term. Here, the function $\sigma(v)$ is the sign function of the object's velocity $v$, i.e., $\sigma(v)=+1,0,-1$ for $v>0,=0,<0$, respectively, and $\Delta_F$ is the coefficient (strength) of the static friction. The $\sigma(v)\Delta_F$ term represents a nonlinearity and, in fact, a singularity in the Langevin equation (\ref{deGennes}), since $\sigma(v)$ is discontinuous at $v=0$. Physically, this term ensures that the solid object is subject to a static friction, which is equal to $\Delta_F$ and acts always, via $\sigma(v)$, opposite to the direction of motion of the object. Both friction coefficients $\alpha$ and $\Delta_F$ are here assumed to be constant, which implies that the object moves over an isotropic surface.

In addition, the motion of the object is driven by externally applied one-dimensional vibrations of the underlying surface, represented by an external noise $\xi(t)$, which has the properties of Gaussian white noise:
\be
\left<\xi(t)\right>&=&0,\\
\label{noise_corr}
\left<\xi(t)\xi(t')\right>&=&2D\delta(t-t'),
\ee
with noise strength $D$. The complicated solid/solid interface is therefore replaced by a static and a dynamic friction term, and randomness is externally induced by Gaussian noise.

In this article we use a path integral approach to study the properties of the nonlinear de Gennes' model Eq.~(\ref{deGennes}). While de Gennes has used a Fokker-Planck approach to obtain approximate results for the transition probability (defined in the next section), the path integral approach provides a more dynamical picture of the statistical properties of the object, on the basis of the most probable or \textit{optimal} paths in the velocity-time plane. Using these optimal paths, we obtain an analytical solution for the transition probability in the saddle-point approximation for small values of $D$. As one of our main results we show that the optimal paths of Eq.~(\ref{deGennes}) can be divided into two classes of paths, which correspond physically to \textit{slip} motion, where the object moves with a velocity $v\neq 0$ over the underlying surface, and \textit{stick-slip} motion, where the object is stuck to the surface with $v=0$ for a finite time. The existence of these two kinds of basic motions is a basic element of almost all dynamical systems with solid/solid friction and appear naturally in de Gennes' model in the path integral framework.

In the following we present a detailed account of the path-integral approach to nonlinear stochastic systems. We analyze the structure of the optimal paths of Eq.~(\ref{deGennes}), and derive an analytical expression for the transition probability, defined in the next section.

\section{Path integral approach}
\label{Sec_pi}

The transition probability or propagator $f(v,t|v_0,t_0)$ gives the probability to find the object with velocity $v$ at time $t$, given that it had a velocity $v_0$ at the initial time $t_0$. Using very many paths generated by the Gaussian white noise $\xi(t)$, the transition probability can be obtained empirically from many realizations of the de Gennes equation~(\ref{deGennes}) for fixed initial and final conditions. In the asymptotic time limit $t\rightarrow \infty$ the transition probability converges to a stationary distribution $p(v)$, which can be derived from Eq.~(\ref{deGennes}). For, introducing an effective potential
\be
\label{U_pot}
U(v)=\frac{v^2}{2\tau_m}+|v|\Delta,
\ee
with the characteristic inertial time $\tau_m= m/\alpha$ and $\Delta\equiv \Delta_F/m$, Eq.~(\ref{deGennes}) takes the form of Brownian motion in the nonlinear potential $U(v)$:
\be
\label{bm}
\dot{v}(t)=-U'(v)+\xi(t)/m.
\ee
Due to the confining character of $U(v)$ a stationary distribution of the velocity coordinate exists and can be calculated from Eq.~(\ref{bm}) using standard methods \cite{Risken}. The result is 
\be
\label{p_stat}
p(v)= Ne^{-\gamma U(v)},
\ee
where $\gamma\equiv m^2/D$ and $N$ is a normalization constant. Clearly, the stationary distribution $p(v)$ is symmetric under a change of sign of $v$. In fact, also the propagator $f(v,t|v_0,t_0)$ has to be symmetric under the change $v_0\rightarrow -v_0$ and $v\rightarrow -v$. This forward/backward symmetry is physically due to the fact that the surface is assumed to be isotropic and the applied noise is symmetric, so that no bias in forward or backward direction is induced. As a consequence, all the statistical properties of the velocity also have this forward/backward symmetry.

In order to obtain the transition probability we use a path integral approach, which was introduced into Statistical Mechanics by Onsager and Machlup \cite{Onsager53,Machlup53} to study the fluctuations of a system in thermal equilibrium. It was generalized to fluctuations in systems in a NESS in Refs. \cite{Taniguchi07,Taniguchi08,Taniguchi08b,Cohen08}. In this approach the transition probability $f(v,t|v_0,t_0)$ is formally expressed as a path integral, i.e., as an integral over all paths leading from the initial state $(v_0,t_0)$ to the final state $(v,t)$. For the dynamics of Eq.~(\ref{deGennes}), this path integral is given by \cite{Feynman}
\begin{eqnarray}
\label{path_int}
f(v,t|v_0,t_0)=\int_{(v_0,t_0)}^{(v,t)} J[v]e^{-\gamma A[\dot{v},v]}\mathcal{D}v,
\end{eqnarray}
where $A[\dot{v},v]$ is a functional of $v(s)$,
\begin{eqnarray}
\label{action}
A[\dot{v},v]=\int_{t_0}^t\mathcal{L}(\dot{v}(s),v(s))\upd s,
\end{eqnarray}
which is usually referred to as the \textit{action} associated with the path $v(s)$. Here, $\mathcal{L}$ is the Lagrangian 
\begin{eqnarray}
\label{lagrangian}
\mathcal{L}(\dot{v},v)=\frac{1}{4}\left(\dot{v}+U'(v)\right)^2.
\end{eqnarray}
In Eq.~(\ref{path_int}) the integral $\int \mathcal{D}v$ denotes an integral over all paths $v(s)$ from $(v_0,t_0)$ to $(v,t)$. The Jacobian $J[v]$ originates from the transformation $\xi(t)\rightarrow v(t)$ and is a functional of $v(s)$ due to the nonlinearity of the force $-U'(v)$ in Eq.~(\ref{bm}) \cite{Graham73,Hunt81}:
\be
\label{jacobian}
J[v]=e^{\frac{1}{2}\int_{t_0}^t U''(v(s))\upd s}.
\ee

We evaluate the path integral Eq.~(\ref{path_int}) in the saddle-point approximation, which proceeds as follows (cf. \cite{Kleinert}). For large $\gamma$ the dominant contribution to the path integral is due to a particular path $v^*(s)$ that maximizes the exponent in Eq.~(\ref{path_int}), or equivalently, which minimizes the action $A[\dot{v},v]$:
\be
\label{variation}
\delta A[\dot{v}^*,v^*]=0.
\ee
This condition yields an Euler-Lagrange (EL) equation
\be
\frac{\upd}{\upd t}\frac{\partial \mathcal{L}}{\partial \dot{v}^*}-\frac{\partial \mathcal{L}}{\partial v^*}=0,
\ee
for the path $v^*(s)$, which is the path with highest probability among all paths connecting $(v_0,t_0)$ and $(v,t)$, i.e., it is the most probable or \textit{optimal} path. We can then expand the action $A[\dot{v},v]$ in the neighborhood of the optimal path using
\be
v(s)=v^*(s)+z(s),
\ee
where $z(t)$ is the deviation from the optimal path. Clearly, the boundary conditions for $z(s)$ are $z(t_0)=z(t)=0$. Expanding the action around $v^*(s)$ yields
\be
\label{expansion}
A[\dot{v},v]&=&A[\dot{v}^*,v^*]+\int_{t_0}^t\upd s \left.\frac{\delta A}{\delta v(s)}\right|_{v^*}z(s)\nonumber\\
&&+\frac{1}{2}\int_{t_0}^t\upd s\int_{t_0}^t\upd s'\left.\frac{\delta^2 A}{\delta v(s)\delta v(s')}\right|_{v^*}z(s)z(s')+...
\ee
Here, the linear term vanishes due to Eq.~(\ref{variation}) and, using Eqs.~(\ref{action}) and (\ref{lagrangian}), the second order term can be calculated as
\be
\frac{1}{2}\int_{t_0}^t\upd s\int_{t_0}^t\upd s'\left.\frac{\delta^2 A}{\delta v(s)\delta v(s')}\right|_{v^*}z(s)z(s')=\int_{t_0}^t\upd s\left[\dot{z}(s)^2+\Omega(v^*(s))z(s)^2\right],
\ee
with
\be
\Omega(v)\equiv U''(v)^2+U'(v)U'''(v).
\ee
The leading orders in the expansion of the action are thus
\be
\label{expansion2}
A[\dot{v},v]&=&A[\dot{v}^*,v^*]+\int_{t_0}^t\upd s\left[\dot{z}(s)^2+\Omega(v^*(s))z(s)^2\right].
\ee
Substituting only the zeroth order term of this expansion into the path integral Eq.~(\ref{path_int}) then yields the saddle-point approximation of the transition probability $f(v,t|v_0,t_0)$:
\be
\label{propagator}
f(v,t|v_0,t_0)\cong J[v^*]e^{-\gamma A[\dot{v}^*,v^*]},
\ee
valid for large $\gamma$. Keeping, in addition, the second order term in Eq.~(\ref{expansion2}) yields the corrected form
\be
f(v,t|v_0,t_0)\cong J[v^*]e^{-\gamma A[\dot{v}^*,v^*]}F[v^*],
\ee
where the fluctuation factor $F[v^*]$ is determined by the path integral
\be
\label{fluc_fac}
F[v^*]&=&\int_{(0,t_0)}^{(0,t)}e^{-\gamma\int_{t_0}^t\upd s\left[\dot{z}(s)^2+\Omega(v^*(s))z(s)^2\right]}\mathcal{D}z.
\ee
The analytic calculation of the fluctuation factor for the nonlinear potential Eq.~(\ref{U_pot}) is beyond the scope of this article. In the following we focus on the properties of the optimal paths and neglect the second order term in the expansion Eq.~(\ref{expansion2}).

\section{Solution of the Euler-Lagrange equation}

For the Lagrangian Eq.~(\ref{lagrangian}) the EL-equation assumes the explicit form
\be
\label{EL}
\ddot{v}^*-\frac{v^*}{\tau_m^2}-\sigma(v^*)\frac{\Delta}{\tau_m}=0.
\ee
A complete picture of the properties of the optimal paths and, on the basis of Eq.~(\ref{propagator}), of the transition probability $f(v,t|v_0,t_0)$ is obtained by solving Eq.~(\ref{EL}) under the given boundary conditions, which are fixed initial and final velocities $(v_0,t_0)$ and $(v,t)$, respectively. Eq~(\ref{EL}) consists in fact of two different equations, namely one for positive $v^*$, in which case $\sigma(v^*)=+1$ and one for negative $v^*$, where $\sigma(v^*)=-1$. Each of these two equations are straightforward to solve, if the boundary conditions are such that $v^*$ is always positive or negative, i.e., if $v^*(s)$ remains entirely on either the upper ($v>0$) or the lower half ($v<0$) of the velocity-time ($v$-$s$) plane. In that case one finds the solutions (dropping the $*$ for optimal path in the following)
\be
\label{sol_basic}
v_\pm(s)=B_\pm e^{s/\tau_m}+C_\pm e^{-s/\tau_m}\mp\Delta\tau_m,
\ee
where the constants $B_\pm$ and $C_\pm$ are determined by the boundary conditions (cf. Appendix~\ref{Apre}) and $+$ refers to the upper half plane and $-$ to the lower, respectively.

The forward/backward symmetry of the velocity statistics (discussed below Eq.~(\ref{p_stat})) implies that the basic solutions $v_+(s)$ and $v_-(s)$ are symmetric with respect to the $v=0$ axis: the path $v_+(s)$ between $(v_0,t_0)$ and $(v,t)$ on the upper half plane is just the mirror image of the path $v_-(s)$ between $(-v_0,t_0)$ and $(-v,t)$ on the lower half plane. Therefore, in the following discussion it is sufficient to consider only a positive initial velocity $v_0$, without loss of generality. Due to this symmetry, the action for the paths $v_+(s)$ and $v_-(s)$ is the same and can be calculated by substituting Eq.~(\ref{sol_basic}) into Eqs.~(\ref{lagrangian}) and~(\ref{action}). This yields for the basic action
\be
\label{a_basic}
A[\dot{v}_\pm,v_\pm]=\int_{t_0}^t\mathcal{L}(\dot{v}_\pm,v_\pm) \upd s=\Lambda(v,t;v_0,t_0),
\ee
where we define
\be
\label{lambda}
\Lambda(v,t;v_0,t_0)\equiv\frac{\left(e^{t/\tau_m}(\Delta\tau_m+|v|)-e^{t_0/\tau_m}(\Delta\tau_m+v_0)\right)^2}{2\tau_m\left(e^{2t/\tau_m}-e^{2t_0/\tau_m}\right)}.
\ee
In addition to the set of basic solutions of Eq.~(\ref{sol_basic}), another formal solution of the EL-equation~(\ref{EL}) is given by $v(s)=0$. Using the formal solutions $v_\pm(s)$ and $v(s)=0$ one can construct the full solution of the EL-equation for fixed initial and final points as linear combinations of the three formal solutions. Two distinct classes of solutions then arise, namely \textit{indirect} paths that partly follow the $v=0$ axis, and \textit{direct} paths that do not. They are both discussed in the following.

\subsection{Direct paths}

\begin{figure}
\begin{center}
\includegraphics[width=10cm]{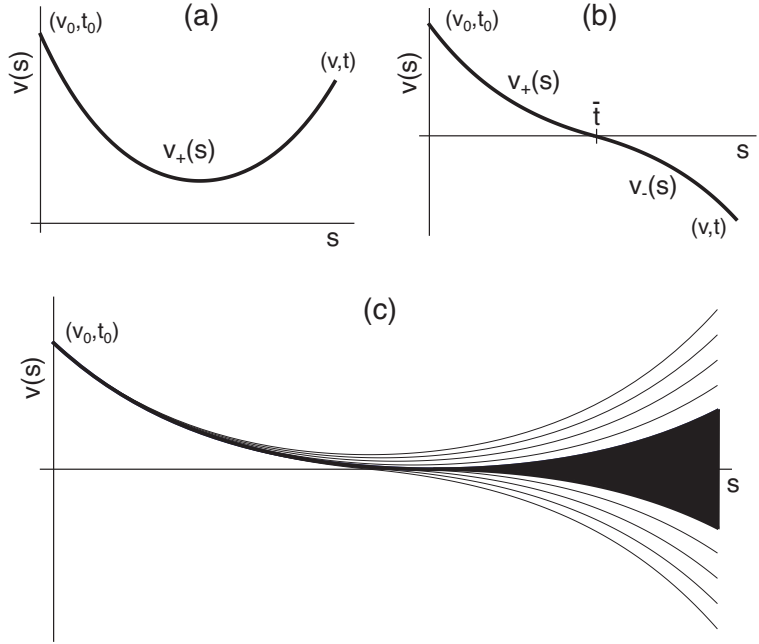}
\caption{\label{direct}Direct paths in the $v$-$s$-plane. (a) A direct path on the upper half plane, parametrized by Eq.~(\ref{dir_par}). (b) A direct crossing path, parametrized by Eq.~(\ref{dirx_par}). (c) Here, we plot a number of direct paths that are very close together initially. When a direct path crosses the $v=0$ axis a jump from a positive to a negative curvature occurs, while direct paths that remain on the upper half plane continue with a positive curvature. The jump in the curvature leads to a region that can not be reached by any direct path from a given initial point $(v_0,t_0)$ (black region).}
\end{center}
\end{figure}

\textit{Direct} paths are characterized by continuous $v(s)$ and $\dot{v}(s)$. They remain either entirely on one half of the $v$-$s$ plane or cross the $v=0$ axis. The former are given by the solutions $v_\pm(s)$, while paths that cross the $v=0$ axis consist of one branch on the upper half plane and one on the lower half plane (cf. Fig.~\ref{direct}(a) and (b)). The direct paths (indicated by the subscript $d$) that remain on the upper plane are simply parametrized by
\be
\label{dir_par}
\left.v_d(s)\right|_{v_0,t_0}^{v,t}=\left.v_+(s)\right|_{v_0,t_0}^{v,t},
\ee
where $v_+(s)$ is given by Eq.~(\ref{sol_basic}) and the boundary conditions are indicated.

For the direct crossing paths we have to consider that the upper branch is given by $v_+(s)$ under the boundary conditions $(v_0,t_0)$ and $(0,\bar{t})$, where $\bar{t}$ is the time at which the path crosses the $v=0$ axis, and the lower branch is given by $v_-(s)$ under the boundary conditions $(0,\bar{t})$ and $(v,t)$. Direct crossing paths (indicated by the subscript $d$ and the superscript $\times$) are thus parametrized by
\be
\label{dirx_par}
v_d^\times(s)=\left\{\begin{array}{l} \left.v_+(s)\right|_{v_0,t_0}^{0,\bar{t}}\quad,\quad t_0\le s\le \bar{t}\\ \\ \left.v_-(s)\right|_{0,\bar{t}}^{v,t}\qquad,\quad \bar{t}< s \le t.
\end{array}\right.
\ee
Here, the crossing time $\bar{t}$ is always $t_0<\bar{t}<t$ and is determined from the condition of a continuous acceleration at the crossover point:
\be
\dot{v}_+(\bar{t})=\dot{v}_-(\bar{t}).
\ee
Using Eq.~(\ref{sol_basic}), this condition leads to a fourth order equation for $\bar{t}$:
\be
\label{tbar_eq}
0&=& \mu^4\Delta\tau_m-\mu^3\left[(\Delta\tau_m+v_0)e^{t_0/\tau_m}+(\Delta\tau_m-v)e^{t/\tau_m}\right]\nonumber\\
&&+\mu\left[(\Delta\tau_m+v_0)e^{(t_0+2t)/\tau_m}+(\Delta\tau_m-v)e^{(2t_0+t)/\tau_m}\right]\nonumber\\
&&-\Delta\tau_me^{(2t_0+2t)/\tau_m},
\ee
where $\mu= e^{\bar{t}/\tau_m}$. Eq.~(\ref{tbar_eq}) has a unique real root $t_0<\bar{t}<t$.

The action associated with the direct paths that remain on the upper half plane is
\be
\label{Adir}
A[\dot{v}_d,v_d]&=&A[\dot{v}_+,v_+]=\Lambda(v,t;v_0,t_0),
\ee
which follows immediately from Eqs.~(\ref{dir_par}) and (\ref{a_basic}). The action of the direct crossing paths on the other hand, consists of contributions from the upper and the lower branch, i.e.,
\be
\label{Adir_x}
A[\dot{v}^\times_d,v^\times_d]&=&\int_{t_0}^{\bar{t}}\mathcal{L}(\dot{v}_+,v_+) \upd s+\int_{\bar{t}}^t \mathcal{L}(\dot{v}_-,v_-) \upd s\nonumber\\
&=&\Lambda(0,\bar{t};v_0,t_0)+\Lambda(v,t;0,\bar{t}),
\ee
using Eqs.~(\ref{dirx_par}) and (\ref{a_basic}). Throughout this paper, the function $\Lambda$ of Eq.~(\ref{lambda}) is adapted to the case at hand, by replacing $v, t;v_0,t_0$ by the appropriate velocities and times.

A crucial observation is then that not all initial and final points in the entire velocity-time plane can be connected by a direct path (cf. Fig.~\ref{direct}(c)). This is due to a jump in the curvature of the direct path, when the $v=0$ axis is crossed: the upper branch $v_+(s)$ is always convex with $\ddot{v}_+\approx\Delta$ close to the $v=0$ axis, while the lower branch $v_-(s)$ is always concave with $\ddot{v}_-\approx-\Delta$ close to the $v=0$ axis (cf. Eq.~(\ref{EL})). There exists thus a region in the $v$-$s$-plane that cannot be reached by any direct path from a given initial point. As $t-t_0$ becomes larger this region grows exponentially, so that eventually, as $t\rightarrow \infty$, fewer and fewer final points can be reached by a direct optimal path. However, this seems to contradict the fact that a stationary distribution exists and has to be approached by $f(v,t|v_0,t_0)$ in the asymptotic time limit. Therefore, direct optimal paths can not represent the full solution of the EL-equation~(\ref{EL}). The key is to consider other solutions that satisfy Eq.~(\ref{EL}) piecewise. This allows us to construct another class of solutions, namely indirect paths.

\subsection{Indirect paths}

\begin{figure}
\begin{center}
\includegraphics[width=10cm]{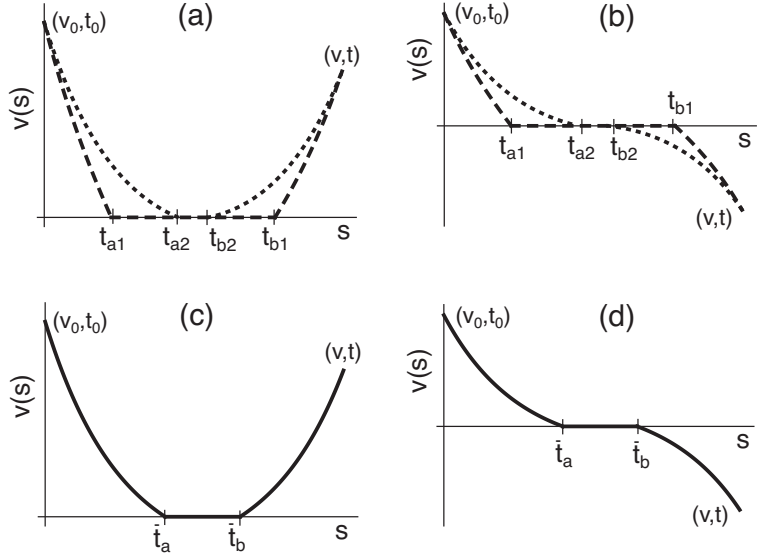}
\caption{\label{indirect}Indirect paths in the $v$-$s$-plane. (a) Two examples of indirect paths on the upper half plane, parametrized by Eq.~(\ref{id_par}). Path $1$ (long dashed curve) crosses the $v=0$ axis at the times $t_{a1}$ and $t_{b1}$. Path $2$ (short dashed curve) at the times $t_{a2}$ and $t_{b2}$. (b) Two indirect paths that cross the $v=0$ axis, parametrized by Eq.~(\ref{id_par}). (c) and (d): The unique indirect paths with minimal action between given initial and final points, parametrized by Eq.~(\ref{idb_par}).}
\end{center}
\end{figure}

\textit{Indirect} paths consist of three parts: a relaxation branch from the initial point $(v_0,t_0)$ to the axis at $(0,t_a)$, a part along the zero axis from $(0,t_a)$ to $(0,t_b)$, and an excitation branch from $(0,t_b)$ to the final point $(v,t)$. The relaxation branch is given by $v_+(s)$ under the boundary conditions $(v_0,t_0)$ and $(0,t_a)$ and the excitation branch by $v_+(s)$ or $v_-(s)$ under the boundary conditions $(0,t_b)$ and $(v,t)$ (cf. Fig.~\ref{indirect}(a) and (b)). Indirect paths (indicated by the subscript $id$) are thus parametrized by
\be
\label{id_par}
v_{id}(s)=\left\{\begin{array}{l} \left.v_+(s)\right|_{v_0,t_0}^{0,t_a}\quad,\quad t_0\le s\le t_a\\ \\
0\quad\qquad\qquad,\quad t_a<s<t_b\\ \\
\left.v_\pm(s)\right|_{0,t_b}^{v,t}\;\,\quad,\quad t_b\le s \le t.
\end{array}\right.
\ee
Clearly, the times $t_a$, $t_b$ have to satisfy the conditions $t_0< t_a\le t_b<t$. All three parts of $v_{id}(s)$ satisfy the EL-equation piecewise. We note that indirect paths are valid solutions of Eq.~(\ref{EL}) because in the path integral formalism the boundary conditions are the fixed initial and final points of the optimal path. If one specifies instead the initial velocity and the initial acceleration of the object, indirect paths can not arise.

The action associated with these indirect paths is
\be
\label{Aind}
A[\dot{v}_{id},v_{id}]&=&\int_{t_0}^{t_a}\mathcal{L}(\dot{v}_+,v_+) \upd s+\int_{t_b}^t \mathcal{L}(\dot{v}_\pm,v_\pm) \upd s\nonumber\\
&=&\Lambda(0,t_a;v_0,t_0)+\Lambda(v,t;0,t_b),
\ee
which follows from Eqs.~(\ref{id_par}) and (\ref{a_basic}). Since $t_a$ and $t_b$ are not specified, there are in principle infinitely many indirect paths between $(v_0,t_0)$ and $(v,t)$ possible, which all have different actions according to Eq.~(\ref{Aind}) (cf. Fig.~\ref{indirect}(a) and (b)). But, among all these indirect paths there exists a unique indirect path with minimal action. This ``optimal" indirect path is obtained by determining the minimum of $A[\dot{v}_{id},v_{id}]$ of Eq.~(\ref{Aind}), with respect to $t_a$ and $t_b$. This is a minimization problem subject to the inequality constraint
\be
\label{ineq}
t_a\le t_b.
\ee

In order to find the solution for this minimization we note that
\be
\label{Aind_der1}
\frac{\partial }{\partial t_a}A[\dot{v}_{id},v_{id}]=\frac{\partial }{\partial t_a}\Lambda(0,t_a;v_0,t_0)
\ee
is independent of $t_b$ and has only one zero for $t_a\in [t_0,t]$, at which the axis is crossed with a positive slope. Likewise
\be
\label{Aind_der2}
\frac{\partial }{\partial t_b}A[\dot{v}_{id},v_{id}]=\frac{\partial }{\partial t_b}\Lambda(v,t;0,t_b),
\ee
independent of $t_a$. These properties imply that $A[\dot{v}_{id},v_{id}]$ of Eq.~(\ref{Aind}) has a unique global minimum as a function of $t_a$ and $t_b$ and is monotonically increasing away from it. Due to this monotonicity, the minimum of $A[\dot{v}_{id},v_{id}]$, subject to the inequality constraint $t_a\le t_b$, is either given by the global unconstrained minimum or, if this minimum can not be attained because it violates the inequality constraint Eq~(\ref{ineq}), by a minimum of $A[\dot{v}_{id},v_{id}]$ subject to the equality constraint $t_a=t_b$.

We then obtain for the minimum of $A[\dot{v}_{id},v_{id}]$:\\
(i) The unconstrained minimum is determined by setting each of the time derivatives, Eqs.~(\ref{Aind_der1}) and (\ref{Aind_der2}), equal to zero. This yields the times
\be
\label{tbar1}
\bar{t}_a&\equiv&t_0+\tau_m\ln\left(1+\frac{v_0}{\Delta \tau_m}\right),\\
\label{tbar2}
\bar{t}_b&\equiv&t-\tau_m\ln\left(1+\frac{|v|}{\Delta \tau_m}\right).
\ee
We note that the time $\bar{t}_a$ is just the time at which a noise-free (or average) path, described by Eq.~(\ref{bm}) with $\xi(t)=0$, would relax to the $v=0$ axis starting from $(v_0,t_0)$. Likewise, $\bar{t}_b$ is the time at which a noise-free path starting at $(v,t)$ would reach the axis, moving backward in time. We parametrize the optimal indirect path specified by $\bar{t}_a$ and $\bar{t}_b$ by (cf. Fig.~\ref{indirect}(c) and (d)):
\be
\label{idb_par}
\bar{v}_{id}(s)=\left\{\begin{array}{l} \left.v_+(s)\right|_{v_0,t_0}^{0,\bar{t}_a}\quad,\quad t_0\le s\le \bar{t}_a\\ \\
0\qquad\qquad\quad,\quad \bar{t}_a<s<\bar{t}_b\\ \\
\left.v_\pm(s)\right|_{0,\bar{t}_b}^{v,t}\;\,\quad,\quad \bar{t}_b\le s \le t.
\end{array}\right.
\ee
The associated action is then given by Eq.~(\ref{Aind}), where $t_a$ and $t_b$ are replaced by $\bar{t}_a$ and $\bar{t}_b$, respectively,
\be
\label{Aind2}
A[\dot{\bar{v}}_{id},\bar{v}_{id}]&=&\Lambda(0,\bar{t}_a;v_0,t_0)+\Lambda(v,t;0,\bar{t}_b)\nonumber\\
&=&\Lambda(v,t;0,\bar{t}_b)\nonumber\\
&=&U(v).
\ee
The term $\Lambda(0,\bar{t}_a;v_0,t_0)$ on the right hand side of the first line of Eq.~(\ref{Aind2}) vanishes, because it is the action associated with the noise-free relaxation branch and for these paths the Lagrangian is identically zero (cf. Eq.~(\ref{lagrangian})). The third line follows upon substituting Eq.~(\ref{tbar2}) into Eq.~(\ref{lambda}).

(ii) If the global minimum does not exist, i.e., if $\bar{t}_b<\bar{t}_a$, one has to determine the minimum of Eq.~(\ref{Aind}) under the equality constraint $t_a=t_b$. For this, one has to solve
\be
\frac{\partial }{\partial t_a}\left[\Lambda(0,t_a;v_0,t_0)+\Lambda(v,t;0,t_a)\right]=0,
\ee
for $t_a$. One finds that there is a unique real solution of this equation $\in[t_0,t]$, which is identical with the time $\bar{t}$ determined by Eq.~(\ref{tbar_eq}). The associated action is then given by Eq.~(\ref{Aind}) with $t_a=t_b=\bar{t}$, i.e., $\Lambda(0,\bar{t};v_0,t_0)+\Lambda(v,t;0,\bar{t})$, which is just equal to the action of the direct crossing paths, $A[\dot{v}^\times_d,v^\times_d]$ of Eq.~(\ref{Adir_x}).

\subsection{Optimal paths in the $v$-$s$-plane}
\label{Sec_structure}

Having determined the two classes of solutions of the EL-equation (\ref{EL}), one can now find the unique optimal path between given initial and final points, direct \textit{or} indirect, using the minimal action principle. To this end we compare the actions of the direct and indirect paths separately for the paths on the upper half-plane and for the crossing paths.

\begin{enumerate}
\item{We consider the paths on the upper half-plane: The direct paths are given by Eq.~(\ref{dir_par}) with the associated action $A[\dot{v}_d,v_d]$, Eq.~(\ref{Adir}). The indirect paths are specified by Eq.~(\ref{idb_par}) with the associated action $A[\dot{\bar{v}}_{id},\bar{v}_{id}]$, Eq.~(\ref{Aind2}). From the condition
\be
\label{cond1}
A[\dot{v}_d,v_d]=A[\dot{\bar{v}}_{id},\bar{v}_{id}],
\ee
which, using Eqs.~(\ref{Adir}) and (\ref{Aind2}), is equivalent to
\be
\label{cond1b}
U(v)=\Lambda(v,t;v_0,t_0),
\ee
one can derive a critical value of $v$, denoted by $u^+(t)$, such that $A[\dot{\bar{v}}_{id},\bar{v}_{id}]\le A[\dot{v}_d,v_d]$ if $v\le u^+(t)$. Eq.~(\ref{cond1b}) leads to a quadratic equation for $v$, which has the relevant root
\be
\label{vc_a}
u^+(t)&\equiv& e^{(t-t_0)/\tau_m}(\Delta\tau_m+v_0)-\Delta\tau_m\nonumber\\
&&-\sqrt{\left(v_0^2+2\Delta\tau_m v_0\right)\left(e^{2(t-t_0)/\tau_m}-1\right)}.
\ee
This means that if $v<u^+(t)$ and $t\ge \bar{t}_a$, the action of the indirect path is lower than that of the direct path. The condition $t\ge \bar{t}_a$ is necessary for indirect paths to exist.}

\item{Let us now consider crossing paths. The direct crossing paths $v^\times_d(s)$ are given by Eq.~(\ref{dirx_par}) with the associated action $A[\dot{v}^\times_d,v^\times_d]$, Eq.~(\ref{Adir_x}). As before, the indirect paths are specified by Eq.~(\ref{idb_par}) with the associated action $A[\dot{\bar{v}}_{id},\bar{v}_{id}]$, Eq.~(\ref{Aind2}). The results of the minimization of Eq.~(\ref{Aind}) show the following: When $\bar{t}_b>\bar{t}_a$ the indirect crossing path always has a lower action than the direct crossing path between the same initial and final points, i.e., $A[\dot{\bar{v}}_{id},\bar{v}_{id}]\le A[\dot{v}^\times_d,v^\times_d]$. On the other hand, when $\bar{t}_b< \bar{t}_a$, the indirect paths of Eq.~(\ref{idb_par}) no longer exist and the direct crossing path then has the lowest action. Therefore, from the condition
\be
\bar{t}_b=\bar{t}_a
\ee
one can derive a critical value of $v$ on the lower half plane, namely
\be
\label{vc_b}
u^-(t)=-\Delta\tau_m\left(\frac{\Delta\tau_m}{\Delta\tau_m+v_0}e^{(t-t_0)/\tau_m}-1\right),
\ee
so that the action of the indirect path is lower than that of the direct path, if $v\ge u^-(t)$ and $t\ge \bar{t}_a$.}
\end{enumerate}

\begin{figure}
\begin{center}
\includegraphics[width=12cm]{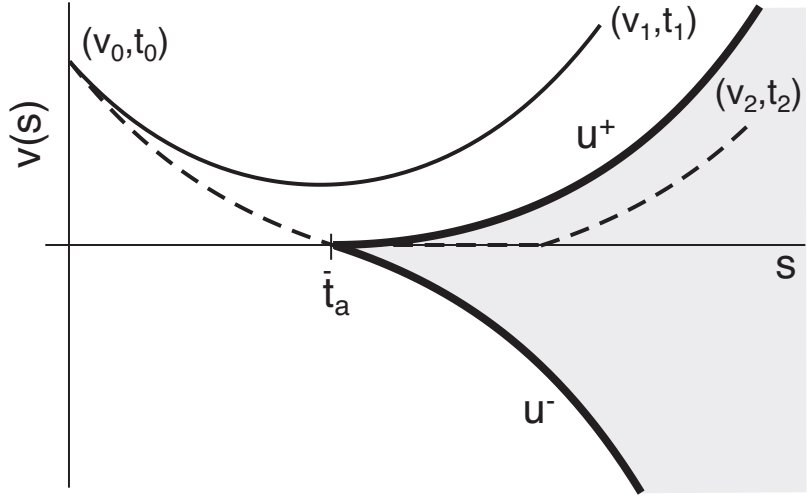}
\caption{\label{phase_dia}Diagram of the optimal paths in a velocity-time plane for a fixed initial point $(v_0,t_0)$ and varying final points. The bold black solid curves are given by $u^+(s)$, Eq.~(\ref{vc_a}), on the upper half plane, and by $u^-(s)$, Eq.~(\ref{vc_b}), on the lower half plane. When the final point of the optimal path lies outside of the shaded region, as here $(v_1,t_1)$, the optimal path is a direct path (solid line). Otherwise, if the final point lies inside the shaded region, as here $(v_2,t_2)$, the optimal path is an indirect path (dashed line).}
\end{center}
\end{figure}

It follows from this discussion that for a given initial point $(v_0,t_0)$ the optimal path is an indirect path if the end point $(v,t)$ lies in the interval $u^-(t)\le v \le u^+(t)$ with $t\ge \bar{t}_a$, otherwise the optimal path is a direct path (cf. Fig.~\ref{phase_dia}).

The curves $u^+(t)$ and $u^-(t)$ represent boundaries in the $v$-$s$ plane, separating different qualitative dynamical behaviors of the moving object in terms of direct and indirect optimal paths. In fact, physically, direct paths can be considered as representing a pure \textit{slip} motion of the object, where $v\neq 0$ apart from one crossing point. Indirect paths follow the $v=0$ axis for a finite time and thus represent physically a \textit{stick-slip} motion.

\section{Transition probability}

From the structure of the optimal paths in the $v$-$s$-plane, the transition probability $f(v,t|v_0,t_0)$ follows directly via the saddle-point approximation Eq.~(\ref{propagator}). The Jacobian as given by Eq.~(\ref{jacobian}) reads explicitly for the potential $U(v)$ of Eq.~(\ref{U_pot}):
\be
\label{jacob}
J[v^*]=e^{(t-t_0)/(2\tau_m)+\Delta\int_{t_0}^t \delta(v^*(s))\upd s}.
\ee
Here, the term $\int_{t_0}^t \delta(v^*(s))\upd s$ is different from zero only when the optimal path $v^*(s)$ crosses the $v=0$ axis. This means that for the direct paths on the upper half plane the Jacobian is just a function of $t_0,t$, which can effectively be absorbed into the normalization of the transition probability. Only for the direct crossing paths and the indirect paths the Jacobian contributes significantly. For the direct crossing paths we obtain
\be
\label{j1}
\int_{t_0}^t \delta\left(v^\times_d(s)\right)\upd s=\frac{1}{|\dot{v}_d^\times(\bar{t})|},
\ee
and for the indirect paths
\be
\label{j2}
\int_{t_0}^t \delta\left(\bar{v}_{id}(s)\right)\upd s=\frac{1}{2|\dot{\bar{v}}_{id}(\bar{t}_a)|}+\frac{1}{2|\dot{\bar{v}}_{id}(\bar{t}_b)|}=\frac{1}{\Delta}.
\ee
In deriving these expressions we have used the representation of the delta function \cite{Arfken}
\be
\delta(g(x))=\sum_i\frac{1}{|g'(x_i)|}\delta(x-x_i),
\ee
where the $x_i$ are the zeros of $g(x)$. In the last step of Eq.~(\ref{j2}) we have substituted $\bar{t}_a$ and $\bar{t}_b$ from Eqs.~(\ref{tbar1}) and (\ref{tbar2}) into the time derivative of $\bar{v}_{id}$, Eq.~(\ref{idb_par}).

With these results for the Jacobian, we can express the transition probability given by Eq.~(\ref{propagator}) as follows. The initial point $(v_0,t_0)$ is given.

For $t\le\bar{t}_a$ (defined in Eq.~(\ref{tbar1})), no indirect optimal paths occur, so that for a final velocity in the interval $-\infty< v\le 0$ only the direct crossing paths contribute to the action in the expression for the transition probability, Eq.~(\ref{propagator}), while for $0<v<\infty$ only direct paths on the upper plane contribute. Using Eqs.~(\ref{Adir}) and (\ref{Adir_x}), respectively, in Eq.~(\ref{propagator}), and considering the contribution of the Jacobian, Eq.~(\ref{jacob}) with Eq.~(\ref{j1}), we obtain the transition probability
\be
\label{sol1}
f(v,t|v_0,t_0)=\mathcal{N}_1\left\{\begin{array}{l} e^{-\gamma \left[\Lambda(0,\bar{t};v_0,t_0)+\Lambda(v,t;0,\bar{t})\right]+\Delta/|\dot{v}_d^\times(\bar{t})|}\quad, \quad-\infty< v\le 0\\ \\ 
e^{-\gamma \Lambda(v,t;v_0,t_0)}\;\;\qquad\qquad\qquad\qquad, \quad 0< v< \infty,
\end{array}\right.
\ee
where $\mathcal{N}_1$ is a normalization factor.

For $t>\bar{t}_a$ indirect paths contribute and the structure of the optimal paths (as discussed in Sec.~\ref{Sec_structure}) indicates that direct crossing paths contribute to the action in Eq.~(\ref{propagator}) for a final velocity in the interval $-\infty< v\le u^-(t)$, while indirect paths contribute for $u^-(t)<0<u^+(t)$, and direct paths on the upper plane for $u^+(t)\le v<\infty$. Using Eqs.~(\ref{Adir}), (\ref{Adir_x}) and (\ref{Aind}), respectively, in Eq.~(\ref{propagator}), and considering the contribution of the Jacobian, Eq.~(\ref{jacob}) with Eqs.~(\ref{j1}) and (\ref{j2}), respectively, the transition probability reads
\be
\label{sol2}
f(v,t|v_0,t_0)=\mathcal{N}_2\left\{\begin{array}{l} e^{-\gamma \left[\Lambda(0,\bar{t};v_0,t_0)+\Lambda(v,t;0,\bar{t})\right]+\Delta/|\dot{v}_d^\times(\bar{t})|}\quad,\quad -\infty< v\le u^-\\ \\
e^{-\gamma U(v)+1}\;\;\quad\qquad\qquad\qquad\qquad,\quad u^-<v<u^+\\ \\
e^{-\gamma \Lambda(v,t;v_0,t_0) }\;\;\qquad\qquad\qquad\qquad, \quad u^+\le v< \infty,
\end{array}\right.
\ee
where $\mathcal{N}_2$ is a normalization factor.

Since both $u^-\rightarrow -\infty$ and $u^+\rightarrow\infty$ as $t\rightarrow\infty$, only indirect paths contribute to the transition probability in the asymptotic time limit and we recover the stationary distribution Eq.~(\ref{p_stat}) in this limit from Eq.~(\ref{sol2}):
\be
\lim_{t\rightarrow\infty}f(v,t|v_0,t_0)=p(v).
\ee

From the transition probability $f(v,t|v_0,t_0)$ one can construct joint probability distributions for arbitrary sequences of $n$-points in the velocity-time plane. Due to the Markovian character of Eq.~(\ref{deGennes}), the joint probability distribution $p(v_n,t_n;...;v_2,t_2;v_1,t_1;v_0,t_0)$, which contains the probability to find the object at the successive points $(v_0,t_0)\rightarrow(v_1,t_1)\rightarrow(v_2,t_2)\rightarrow...\rightarrow(v_n,t_n)$ in the velocity-time plane, is just given by the product of $n$ transition probabilities $f(v_n,t_n|v_{n-1},t_{n-1})$:
\be
\label{joint}
p(v_n,t_n;...;v_2,t_2;v_1,t_1:v_0,t_0)&=&f(v_n,t_n|v_{n-1},t_{n-1})\cdots f(v_2,t_2|v_1,t_1)f(v_1,t_1|v_0,t_0),
\ee
where it is assumed that the object is fixed with the initial velocity $v_0$ at the initial time $t_0$. Associated with the joint probability distribution of Eq.~(\ref{joint}) is then an extended optimal path along the points $(v_0,t_0)\rightarrow(v_1,t_1)\rightarrow(v_2,t_2)\rightarrow...\rightarrow(v_n,t_n)$, which is determined from the structure of the optimal paths in the $v$-$s$-plane, as discussed in Sec.~\ref{Sec_structure}. A segment of the extended optimal path, between two neighboring points $(v_n,t_n)$ and $(v_{n-1},t_{n-1})$, is either a direct or an indirect path depending on the relative location of these two points. 

\section{Conclusion}

We have derived a complete characterization of the optimal paths of the de Gennes' equation~(\ref{deGennes}) within the path integral framework. The optimal paths can be divided into two classes: a) \textit{Direct} optimal paths, with continuous $v(s)$ and $\dot{v}(s)$, which physically can be considered as representing a pure slip motion of the object. b) \textit{Indirect} optimal paths, with continuous $v(s)$ and discontinuous $\dot{v}(s)$, which follow partly the $v=0$ axis and represent physically a stick-slip motion. We have shown that for a given initial point $(v_0,t_0)$ the optimal path will \textit{either} be a direct \textit{or} an indirect path depending on the location of the final point $(v,t)$ in the velocity-time plane (cf. Fig.~\ref{phase_dia}). In the asymptotic time limit $t\rightarrow\infty$ a finite final velocity $v$ can only be reached by an indirect path. 

This analysis of the optimal paths leads to an analytical result for the transition probability $f(v,t|v_0,t_0)$ in the saddle-point approximation. The calculation of correction terms to this result, such as the fluctuation factor $F(v^*)$, Eq.~(\ref{fluc_fac}), which takes into account the second order term in the expansion of the action, Eq.~(\ref{expansion}), is left for future work. However, we want to emphasize that higher order corrections leave the properties and the structure of the optimal paths, and therefore also their slip and stick-slip character, unchanged.

From a physical point of view, the friction terms in Eq.~(\ref{deGennes}) represent a very simple phenomenological model of the solid/solid interaction between the object and the surface. By studying generalizations of Eq.~(\ref{deGennes}), incorporating, e.g., two dimensional or memory effects, one could model more complicated surface properties, such as surface anisotropies or defects. A comparison of such more realistic, but still phenomenological models, with experiments will lead to a better understanding of the effects of surface inhomogeneities.

\begin{appendix}

\section{The prefactors $B_\pm$ and $C_\pm$}
\label{Apre}

The basic solutions $v_+(s)$ and $v_-(s)$ are given by Eq.~(\ref{sol_basic}), where $+$ refers to the upper half of the $v$-$s$-plane and $-$ to the lower half, respectively. The prefactors $B_+$ and $C_+$, for the upper half plane, are determined by the boundary conditions, i.e., by the conditions that the path is initially at $(v_0,t_0)$ and ends at $(v,t)$:
\be
v_+(t_0)=v_0, \qquad v_+(t)=v.
\ee
Solving Eq.~(\ref{sol_basic}) under these boundary conditions for $B_+$ and $C_+$ yields
\be
\label{B+}
B_+&=&\frac{e^{t/\tau_m}(v+\Delta \tau_m)-e^{t_0/\tau_m}(v_0+\Delta \tau_m)}{e^{2t/\tau_m}-e^{2t_0/\tau_m}},\\
\label{C+}
C_+&=&\frac{e^{t/\tau_m}(v_0+\Delta \tau_m)-e^{t_0/\tau_m}(v+\Delta \tau_m)}{e^{(t-t_0)/\tau_m}-e^{-(t-t_0)/\tau_m}}.
\ee
The prefactors $B_-$ and $C_-$, for the lower half plane, are then obtained by simply changing $\Delta \rightarrow -\Delta$ in Eqs.~(\ref{B+}) and (\ref{C+}).

\end{appendix}


\begin{thebibliography}{12}
\expandafter\ifx\csname natexlab\endcsname\relax\def\natexlab#1{#1}\fi
\expandafter\ifx\csname bibnamefont\endcsname\relax
  \def\bibnamefont#1{#1}\fi
\expandafter\ifx\csname bibfnamefont\endcsname\relax
  \def\bibfnamefont#1{#1}\fi
\expandafter\ifx\csname citenamefont\endcsname\relax
  \def\citenamefont#1{#1}\fi
\expandafter\ifx\csname url\endcsname\relax
  \def\url#1{\texttt{#1}}\fi
\expandafter\ifx\csname urlprefix\endcsname\relax\def\urlprefix{URL }\fi
\providecommand{\bibinfo}[2]{#2}
\providecommand{\eprint}[2][]{\url{#2}}

\bibitem[{\citenamefont{Persson}(2000)}]{Persson}
\bibinfo{author}{\bibfnamefont{B.~N.~J.} \bibnamefont{Persson}},
  \emph{\bibinfo{title}{Sliding Friction}} (\bibinfo{publisher}{Springer},
  \bibinfo{year}{2000}).

\bibitem[{\citenamefont{Gennes}(2005/06/01/)}]{DeGennes05}
\bibinfo{author}{\bibfnamefont{P.~G.} \bibnamefont{de Gennes}},
  \bibinfo{journal}{J. Stat. Phys.}
  \textbf{\bibinfo{volume}{119}}, \bibinfo{pages}{953}
  (\bibinfo{year}{2005}).

\bibitem[{\citenamefont{Buguin et~al.}(2006)\citenamefont{Buguin, Brochard, and
  de~Gennes}}]{Buguin06}
\bibinfo{author}{\bibfnamefont{A.}~\bibnamefont{Buguin}},
  \bibinfo{author}{\bibfnamefont{F.}~\bibnamefont{Brochard}}, \bibnamefont{and}
  \bibinfo{author}{\bibfnamefont{P.-G.} \bibnamefont{de~Gennes}},
  \bibinfo{journal}{Eur. Phys. J. E} \textbf{\bibinfo{volume}{19}},
  \bibinfo{pages}{31} (\bibinfo{year}{2006}).

\bibitem{Risken}H. Risken, \textit{The Fokker-Planck Equation: Methods of Solution and Applications} (Springer, Berlin, 1996).

\bibitem[{\citenamefont{Feynman and Hibbs}(1965)}]{Feynman}
\bibinfo{author}{\bibfnamefont{R.~P.} \bibnamefont{Feynman}} \bibnamefont{and}
  \bibinfo{author}{\bibfnamefont{A.~R.} \bibnamefont{Hibbs}},
  \emph{\bibinfo{title}{{Quantum Mechanics and Path Integrals}}}
  (\bibinfo{publisher}{McGraw-Hill, New York}, \bibinfo{year}{1965}).

\bibitem[{\citenamefont{Onsager and Machlup}(1953)}]{Onsager53}
\bibinfo{author}{\bibfnamefont{L.}~\bibnamefont{Onsager}} \bibnamefont{and}
  \bibinfo{author}{\bibfnamefont{S.}~\bibnamefont{Machlup}},
  \bibinfo{journal}{Phys. Rev.} \textbf{\bibinfo{volume}{91}},
  \bibinfo{pages}{1505} (\bibinfo{year}{1953}).

\bibitem[{\citenamefont{Machlup and Onsager}(1953)}]{Machlup53}
\bibinfo{author}{\bibfnamefont{S.}~\bibnamefont{Machlup}} \bibnamefont{and}
  \bibinfo{author}{\bibfnamefont{L.}~\bibnamefont{Onsager}},
  \bibinfo{journal}{Phys. Rev.} \textbf{\bibinfo{volume}{91}},
  \bibinfo{pages}{1512} (\bibinfo{year}{1953}).

\bibitem[{\citenamefont{Taniguchi and Cohen}(2007)}]{Taniguchi07}
\bibinfo{author}{\bibfnamefont{T.}~\bibnamefont{Taniguchi}} \bibnamefont{and}
  \bibinfo{author}{\bibfnamefont{E.~G.~D.} \bibnamefont{Cohen}},
  \bibinfo{journal}{J. Stat. Phys.}
  \textbf{\bibinfo{volume}{126}}, \bibinfo{pages}{1} (\bibinfo{year}{2007}).

\bibitem[{\citenamefont{Taniguchi and Cohen}(2008{\natexlab{a}})}]{Taniguchi08}
\bibinfo{author}{\bibfnamefont{T.}~\bibnamefont{Taniguchi}} \bibnamefont{and}
  \bibinfo{author}{\bibfnamefont{E.~G.~D.} \bibnamefont{Cohen}},
  \bibinfo{journal}{J. Stat. Phys.}
  \textbf{\bibinfo{volume}{130}}, \bibinfo{pages}{1}
  (\bibinfo{year}{2008}{\natexlab{a}}).

\bibitem[{\citenamefont{Taniguchi and
  Cohen}(2008{\natexlab{b}})}]{Taniguchi08b}
\bibinfo{author}{\bibfnamefont{T.}~\bibnamefont{Taniguchi}} \bibnamefont{and}
  \bibinfo{author}{\bibfnamefont{E.~G.~D.} \bibnamefont{Cohen}},
  \bibinfo{journal}{J. Stat. Phys.}
  \textbf{\bibinfo{volume}{130}}, \bibinfo{pages}{633}
  (\bibinfo{year}{2008}{\natexlab{b}}).

\bibitem[{\citenamefont{Cohen}(2008)}]{Cohen08}
\bibinfo{author}{\bibfnamefont{E.~G.~D.} \bibnamefont{Cohen}},
  \bibinfo{journal}{J. Stat. Mech.}
  \textbf{\bibinfo{volume}{2008}}, \bibinfo{pages}{P07014}
  (\bibinfo{year}{2008}).

\bibitem[{\citenamefont{Graham}(1973)}]{Graham73}
\bibinfo{author}{\bibfnamefont{R.}~\bibnamefont{Graham}},
  \bibinfo{journal}{Springer Tracts in Modern Physics}
  \textbf{\bibinfo{volume}{66}}, \bibinfo{pages}{1} (\bibinfo{year}{1973}).

\bibitem[{\citenamefont{Hunt and Ross}(1981)}]{Hunt81}
\bibinfo{author}{\bibfnamefont{K.~L.~C.} \bibnamefont{Hunt}} \bibnamefont{and}
  \bibinfo{author}{\bibfnamefont{J.}~\bibnamefont{Ross}}, \bibinfo{journal}{The
  J. Chem. Phys.} \textbf{\bibinfo{volume}{75}},
  \bibinfo{pages}{976} (\bibinfo{year}{1981}).
  
\bibitem{Kleinert}H. Kleinert, \textit{Path Integrals in Quantum Mechanics, Statistics, Polymer Physics, and Financial Markets} (World Scientific, 2009).

\bibitem{Arfken}G. B. Arfken and H. J. Weber, \textit{Mathematical Methods for Physicists} (Academic Press, 2005)

\end{thebibliography}
\end{document}